# Effects of Ni Substitutions on the Critical Behaviors in $Nd_{0.6}Sr_{0.4}Mn_{1-x}Ni_xO_3$ Manganite.

M. Kh. Hamad[a*], Y. Maswadeh[b], and Kh. A. Ziq[a]

[a]Physics Department, King Fahd University of Petroleum and Minerals. Dhahran, 31261, Saudi Arabia.

[b]Department of Physics and Science of Advanced Materials Program, Central Michigan University, Mt. Pleasant, Michigan 48859, United States.

**ABSTRACT**

We investigate the effect of Ni-substitution on the crystalline structure and the critical behavior of $Nd_{0.6}Sr_{0.4}Mn_{1-x}Ni_xO_3$ *(0.00 ≤ x ≤ 0.20)* perovskite. X-ray diffraction patterns revealed that the major phase in all samples is the orthorhombic structure with space group *Pnma*. Rietveld refinement revealed a linear reduction in the lattice parameters along with monotonic reduction in the *O2-Mn-O2* angel with increasing Ni concentration. The modified Arrott plots and the Kouvel-Fisher method have been used to analyze the magnetization isotherms near the paramagnetic to ferromagnetic (PM-FM) phase transition. The obtained critical exponents (β, γ and δ) revealed that the *Ni*-free sample is consistent with 3D-Heisenberg like behavior. However, upon Ni-substitution, the critical exponents exhibit a mean field like behavior. The reliability of the obtained critical exponent (β, γ, and δ) values have been confirmed by the universal scaling behavior of the isothermal magnetization near the transition temperature.

1.  **INTRODUCTION**

The rich magnetic phase diagram often found in manganites arises from the interplay between the spin, orbital, charge, and lattice degrees of freedom. This is reflected in the wide variety in their properties, such as transport, magnetic, magnetoresistance (MR) and magnetocaloric effect (MCE) [1-4].

Perovskite manganite are generally represented as $R_{1-x}A_xMnO_3$, where $R$ is a rare-earth element such as $Nd, La, and\ Ce$, and $A$ is a divalent alkaline earth metal like $Sr, Ca, Ba, and\ etc$. These materials undergo a combination of several phase: metallic, insulating, and various magnetic phase; ferromagnetic, ferrimagnetic, antiferromagnetic and spin-glass [5]. The level of substitution with a divalent alkaline earth metal in these manganites affects the exchange interaction, revealing a wide variation in many physical properties like the magnetoresistance (MR) and magnetocaloric effect (MCE) [6]. Moreover, the paramagnetic to ferromagnetic phase transition and the corresponding critical exponents near the transition temperature is expected to be affected by these substitutions.

The critical exponents are commonly evaluated using the isothermal magnetization along with Kouvel-Fisher procedure [7, 9-10]. However, for various perovskite manganites there is a wide discrepancy in the reported values of the critical exponents covering the range between the short-range Heisenberg model (β = 0.365, γ =1.336, and δ = 4.800) to the long-range mean-field theory (β = 0.5, γ = 1, and δ = 3). These variations may be related to differences in the micro-structure and possible disorder effects caused by variations in the preparation methods [4, 8].

The phase diagram of $Nd_{1-x}Sr_xMnO_3$ perovskite reveals a variety of magneto-electric properties [11]. Near x =0.4, the material is insulating ferromagnet with maximum Curie temperature $T_c$ ~ 300K and maximum saturated magnetization. These properties lend the material attractive in near room temperature magnetocaloric cooling, sensing devices and data storage applications [12-13]. Moreover, the effects of transition metal substitutions on various magnetic states and their critical behavior has received little attention in this system.

In this paper, we investigate the effects of Ni-substitutions on the critical exponents in $Nd_{0.6}Sr_{0.4}Mn_{1-x}Ni_xO_3$ *(0.00 ≤ x ≤ 0.20)* polycrystalline samples near its FM-PM phase transition. The corresponding critical exponents are determined from the modified Arrott plots [9] and are confirmed by Kouvel-Fisher method [10]. The obtained critical exponents revealed that the unsubstituted sample $Nd_{0.6}Sr_{0.4}MnO_3$ is consistent with 3D-Heisenberg like behavior. Upon Ni-substitution, the critical exponents exhibit a mean field like behavior. The obtained critical exponent ($\beta$, $\gamma$, and $\delta$) have been used to generate the universal scaling behavior of the isothermal magnetization near the transition temperature.

## 2. EXPERIMENTAL DETAILS

The polycrystalline $Nd_{0.6}Sr_{0.4}Mn_{1-x}Ni_xO_3$ (0 ≤ x ≤ 0.2) were prepared using a standard solid-state reaction. Stoichiometric ratios of high purity (4N) oxides ($Nd_2O_3$, $SrCO_3$, $MnO_2$, and $NiO$) were used to prepare the samples. The oxides were mixed and grinded for about 30 minutes. The resulting powder was calcinated in the air for 24 hours at 1000ºC. The samples were re-grinded, pressed and re-annealed at 1000ºC and 1300ºC for 10 hours. The last step was repeated one more time and the samples were furnace cooled to room temperature. Powder x-ray diffraction patterns were obtained using Bruker XRD (Phaser D2 $2^{nd}$-G) with Cu K$\alpha$ ($\lambda$=1.54056 Å). The crystalline structure was analyzed in the range 20-80º using Rietveld refinements available in FULLPROF software. VESTA and Diamond software [14-15] were used to determine the atomic distances and angles between *Mn* ion and the nearest oxygen ions neighbors. The analyses included the refined crystal parameters, the secondary phases, and the cationic distributions in the lattice [16].

A homemade *ac*-susceptometer was fitted inside a Janis closed cycle refrigerator (model SHI-4T). The susceptometer was used to monitor the variation of the susceptibility with temperature in the temperature range (4 to 325K). LABVIEW software is used to control the lock-in-amplifier (SR850) and the Lake Shore 336 temperature controller. The Magnetization measurements were performed using a 9-Tesla PAR-Lakeshore (Model 4500/150A) vibrating sample magnetometer (VSM).

## 3. RESULTS AND ANALYSIS

### 3.1 X-ray Diffraction analysis (XRD)

A representative of the XRD patterns for $Nd_{0.6}Sr_{0.4}Mn_{1-x}Ni_xO_3$ for $x = 0.00$ and $x = 0.05$ along with Rietveld refinement [17, 18] are shown in Fig.1.

The results of the refinement revealed that the orthorhombic structure as the major phase in all samples ($Nd_{0.6}Sr_{0.4}Mn_{1-x}Ni_xO_3$) with *Pnma* (62) space group. A minor cubic perovskite phase with space group $Pm\bar{3}m$ *(221)* has also been detected in all samples. Venkatesh *et al*. reported the presence of pseudocubic phase in $Nd_{0.67}Sr_{0.33}MnO_3$ sample and the orthorhombic phase in $Nd_{0.6}Sr_{0.4}MnO_3$; however, no details were given about heat treatment [12]. Abdel-Latif *et al* used chemical co-precipitation method and low annealing temperature (850 ºC) to produce pure orthorhombic phase of $Nd_{0.6}Sr_{0.4}MnO_3$ [19-20]. Moreover, Mansour *et al* used high temperature annealing (1200~1300 ºC) to prepare $Nd_{0.6}Sr_{0.4}MnO_3$, they obtained almost pure orthorhombic phase with traces of the cubic phase [13].

We performed the refined lattice parameters, unit cell volume, and reliability (R) factors for both phases. The results are listed in Table 1 and in figures 2. We report specifically the $R_f$-factor and the Bragg $R_B$-factor; these factors are referring to the crystallographic agreement and the reflected intensities agreement consecutively. The values of these factors (Table 1) indicate a very good agreement between the XRD patterns and the reflected intensities generated by the space groups.

The results (Table 1) shown in Fig. 2 indicates almost linear reduction in the lattice parameters *(a, b, and c)*, and the unit cell volume (V) with increasing Ni concentration. The reduction in the lattice parameters is consistence with the smaller ionic radius of Ni as compared to Mn ionic radius [21-22]. Upon Ni-substitution at the Mn site, the (200) peak (inset to figure 1) shifts towards higher angles. This indicates a gradual reduction in the lattice parameter *(a)* (Table 1) which is in line with the observed shortening in the *Mn-O* bonds (Table 2). Moreover; this is consistent with a slight increase in the concentration of the smaller $Mn^{4+}$ ions (0.530A°) and decrease in the concentration of the larger $Mn^{3+}$ ions (0.645A°) [23]. This is further supported by the slight decrease in the saturated magnetization of sample with higher Ni-concentration -see figure 7.

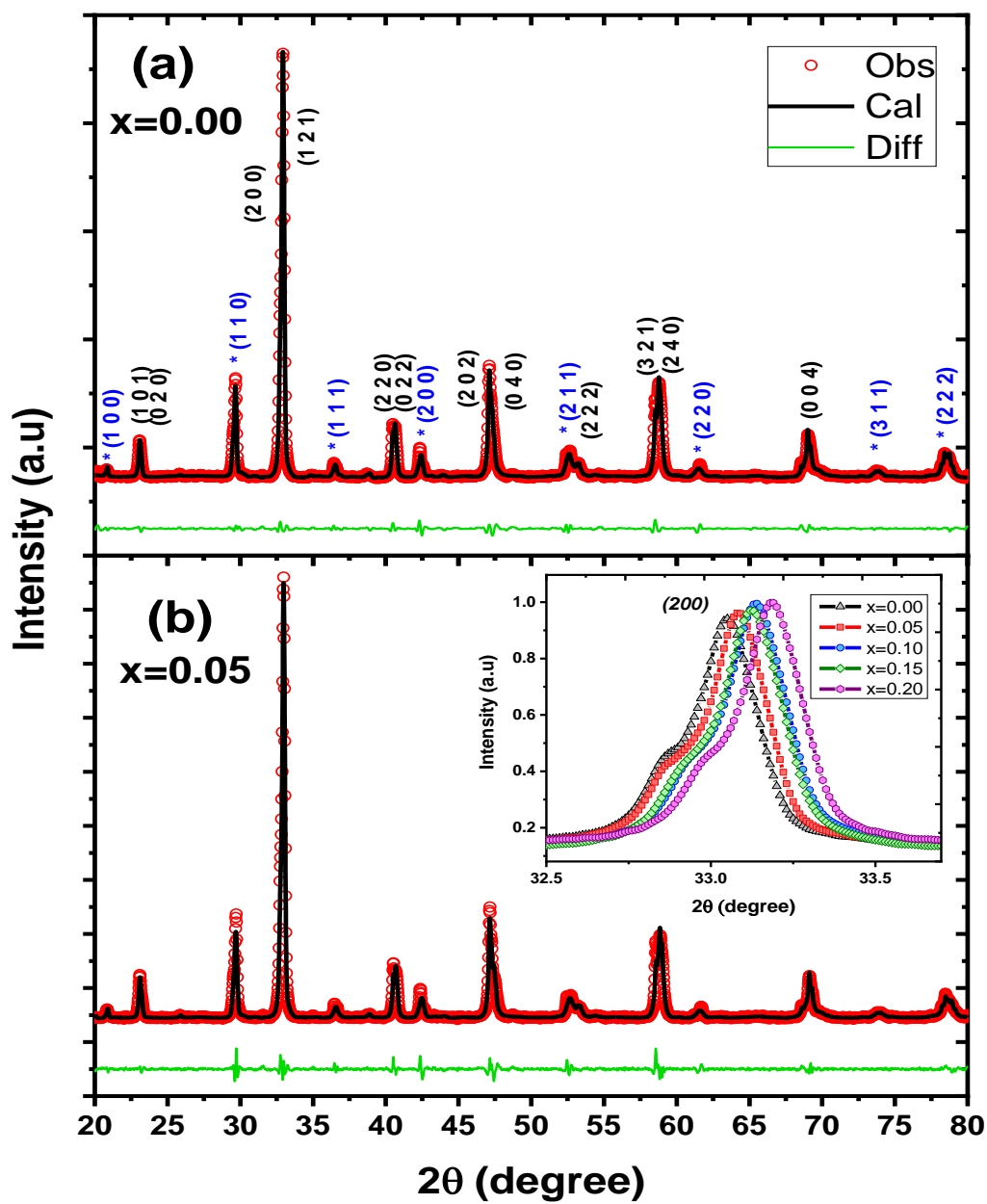

*Figure 1: X-ray diffraction patterns for (a) x = 0.00, and (b) x = 0.05 sample at room temperature. The (hkl) values corresponding to Bragg peaks of primary (in black) and secondary (in blue with stars) phases are marked. The inset shows (200) reflection peaks for different Ni concentration (0.00≤x≤0.20). The diffraction peak is shifted positively in 2θ upon increasing the Ni content in the sample.*

The change in lattice parameters leads to changes in the bond lengths *(Mn-O)*, and hence the bonds angles *(O-Mn-O)* as shown in Table 2 and figure 3. VESTA software was used to determine the distances *(Mn-O)* between *Mn* ions and the nearest neighbors of oxygen ions as well as the *(O-Mn-O)* bond angles. The coordination octahedral of *Mn* is presented in Fig. 4. The Wyckoff positions of ions in the orthorhombic structure are also given (see Table 3).

*Table 1: Summary of the refinement of XRD patterns of $Nd_{0.6}Sr_{0.4}Mn_{1-x}Ni_xO_3$ manganites.*

| Concentration (x) | 0 | 0.05 | 0.10 | 0.15 | 0.20 |
|---|---|---|---|---|---|
| Phase 1: $Nd_{0.6}Sr_{0.4}Mn_{1-x}Ni_xO_3$ | | Space group: Pnma (62) – orthorhombic | | | |
| $a(Å)$ | 5.47(0) | 5.46(6) | 5.46(5) | 5.45(8) | 5.45(3) |
| $b(Å)$ | 7.67(5) | 7.66(0) | 7.66(0) | 7.65(2) | 7.64(7) |
| $c(Å)$ | 5.43(6) | 5.43(1) | 5.42(8) | 5.42(0) | 5.41(4) |
| $V(Å^3)$ | 228.19 | 227.41 | 227.23 | 226.34 | 225.74 |
| $R_B$ | 1.91 | 2.04 | 1.5 | 1.15 | 2.53 |
| $R_F$ | 3.18 | 2.47 | 2.27 | 1.71 | 2.68 |
| Phase 2: $Nd(Sr)MnO_3$ | | Space group: $Pm\bar{3}m$ (221) – cubic | | | |
| $a(Å)$ | 4.26(3) | 4.25(8) | 4.25(7) | 4.25(1) | 4.24(7) |
| $V(Å^3)$ | 77.45 | 77.20 | 77.15 | 76.82 | 76.61 |
| $R_B$ | 2.38 | 1.71 | 2.6 | 1.94 | 1.89 |
| $R_F$ | 1.87 | 1.32 | 1.78 | 1.28 | 1.10 |

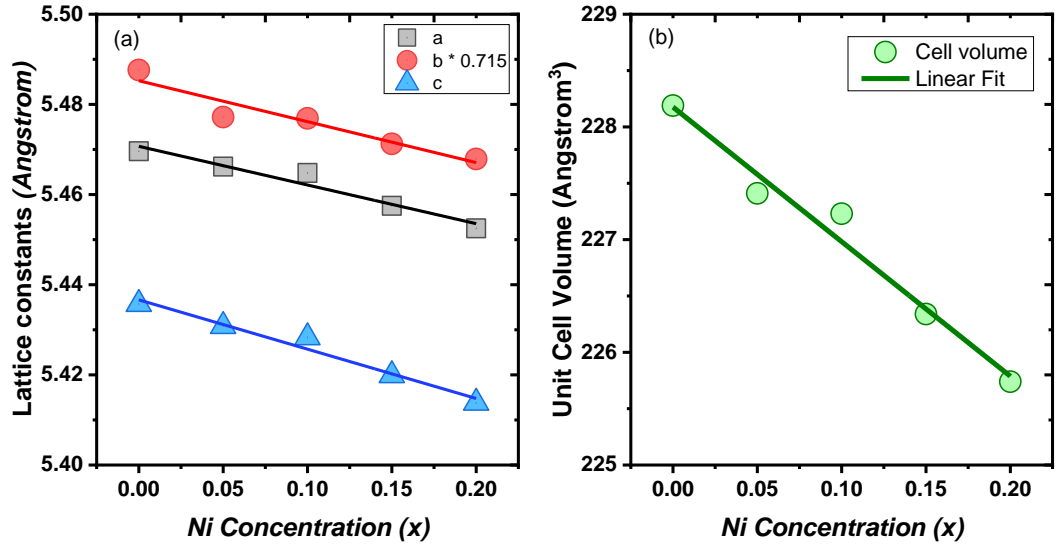

*Figure 2: Change of lattice constants and unit cell volume with increasing the Ni fraction.*

*Table 2: Bonds length and bonds angles of $Nd_{0.6}Sr_{0.4}Mn_{1-x}Ni_xO_3$ manganites.*

| x | Mn-O1 (Å) | Mn-O2 (Å) | Mn-O2 (Å) | O1-Mn-O1 (deg) | O2-Mn-O2 (deg) |
|---|---|---|---|---|---|
| 0.00 | 1.94(8) | 1.95(8) | 1.98(9) | 180 | 90.97(5) |
| 0.05 | 1.93(1) | 1.92(4) | 1.99(3) | 180 | 90.36(7) |
| 0.10 | 1.94(6) | 1.94(9) | 1.96(6) | 180 | 90.05(6) |
| 0.15 | 1.95(2) | 1.93(0) | 1.95(7) | 180 | 89.76(5) |
| 0.20 | 1.92(7) | 1.91(7) | 1.97(2) | 180 | 89.54(0) |

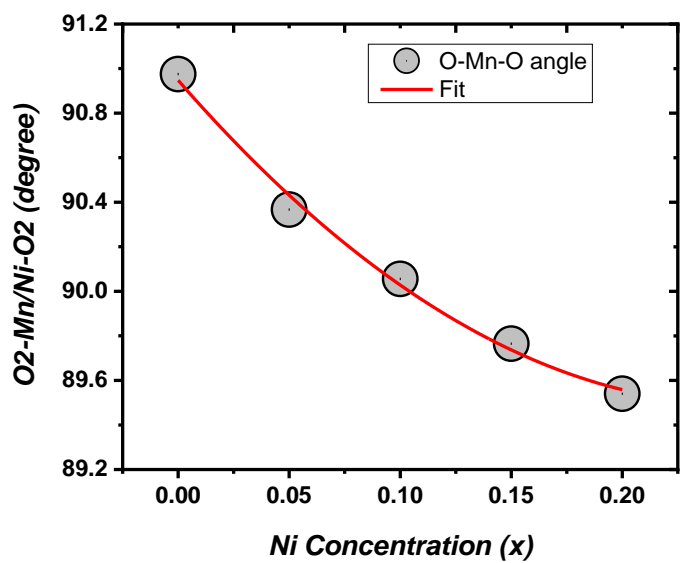

*Figure 3: Change of bond angles of Nd$_{0.6}$Sr$_{0.4}$Mn$_{1-x}$Ni$_x$O$_3$ upon Ni concentration.*

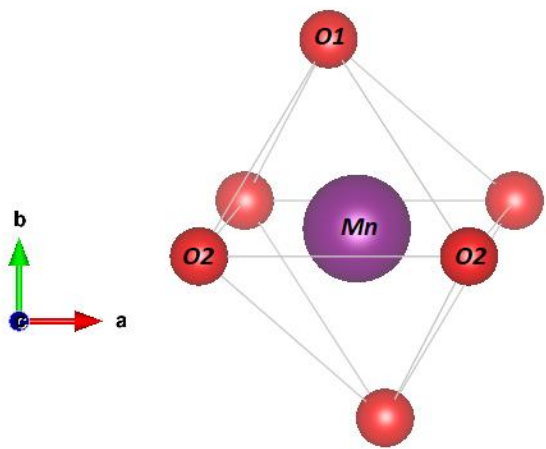

*Figure 4: The coordinate of polyhedral of Mn in Nd$_{0.6}$Sr$_{0.4}$Mn$_{1-x}$Ni$_x$O$_3$ compound.*

*Table 3: Wyckoff positions of ions in the orthorhombic structure for Pnma (62) space group.*

| *Pnma (62) - orthorhombic* | | | | |
|---|---|---|---|---|
| **Atom** | **Wyck.** | **x/a** | **y/b** | **z/c** |
| Nd | 4c | 0.012(5) | 0.250(0) | 0.005(4) |
| Sr | 4c | 0.012(5) | 0.250(0) | 0.005(4) |
| Mn | 4b | 0.0000 | 0.0000 | 0.500(0) |
| O1 | 4c | 0.467(9) | 0.250(0) | 0.031(7) |
| O2 | 8d | 0.281(1) | 0.027(8) | 0.725(7) |
| Ni | 4b | 0.0000 | 0.0000 | 0.500(0) |

Distortion of the ideal perovskite structure is very important parameter since it correlates with magnetic properties. The obtained distortion index *(D)*, quadratic elongation, Octahedrons volume, and bond angle variance are summarized in Table 4. There is no major structural deformation on the $MnO_6$ octahedron upon Ni concentration, the quadratic elongation values are very close to 1 (1.006 ± 0.003), and the calculated distortion index are very small (~0.010 ± 0.005), which indicates that the calculated structure is in good agreement with the structure of the material. The octahedron volume is decreasing with increasing Ni concentration, consistent with the smaller Ni-ionic radius compared with Mn-ionic radius, further supporting that Ni is substituting Mn 4b site.

*Table 4: Octahedral volume, distortion index and quadratic elongation of Mn polyhedral in*

*$Nd_{0.6}Sr_{0.4}Mn_{1-x}Ni_xO_3$*

| X | Octahedral volume (Å³) | Distortion index (bond length) | Quadratic elongation |
|---|---|---|---|
| 0.00 | 9.98(7) | 0.008(1) | 1.00(9) |
| 0.05 | 9.78(7) | 0.015(0) | 1.00(7) |
| 0.10 | 9.85(8) | 0.004(3) | 1.00(6) |
| 0.15 | 9.72(0) | 0.005(7) | 1.00(8) |
| 0.20 | 9.66(5) | 0.011(3) | 1.00(4) |

## 3.2 AC-susceptibility

The ac-susceptibility for all samples are measured at 887Hz and the amplitude of the field $H_{ac}=0.1 Oe$. The normalized real part of the susceptibility is presented in Fig. 5a and its derivative in Fig. 5b. The position of the minimum in the derivative is taken as the Curie temperature. Figure 5b reveals a gradual reduction in the transition temperature along with increase in the full-width at half maximum *(FWHM)* with increasing Ni-concentration. The reduction in Tc is presented in Fig. 6, indication linear decrease in Tc. It is worth mentioning that upon increasing the frequency, there is a very small (less than 0.10 K) shift in the peak temperatures over the frequency range (100-5000 Hz) [7].

In Nd-manganites, $Nd$ -ions as $Nd^{3+}$. However; $Mn$ ions may have several oxidation states mostly: $Mn^{4+}, Mn^{3+}$ and $Mn^{2+}$. Substitution with $Ni^{2+}$ at the $Mn$ site ultimately affects the charge distribution and the overall magnetic state of the Nd-manganites. This may results in a reduction of the ferromagnetic transition temperature (Tc) and the associated FM exchange interaction. It also enhances the antiferromagnetic behavior as expected for an oxide material in which the magnetism is mediated by super exchange through the oxygen [24]. The indirect exchange interaction depends on the degree of overlap of orbitals, which in turn is strongly dependent upon the angle of the $Mn - O - Mn$ bond, where the $3d$ electrons delocalized over this bond. This bond is strongly depending upon substitution where $Mn^{4+}/Mn^{3+}$ ratio is affected, and hence interruption of $Mn - O - Mn$ double exchange interaction.

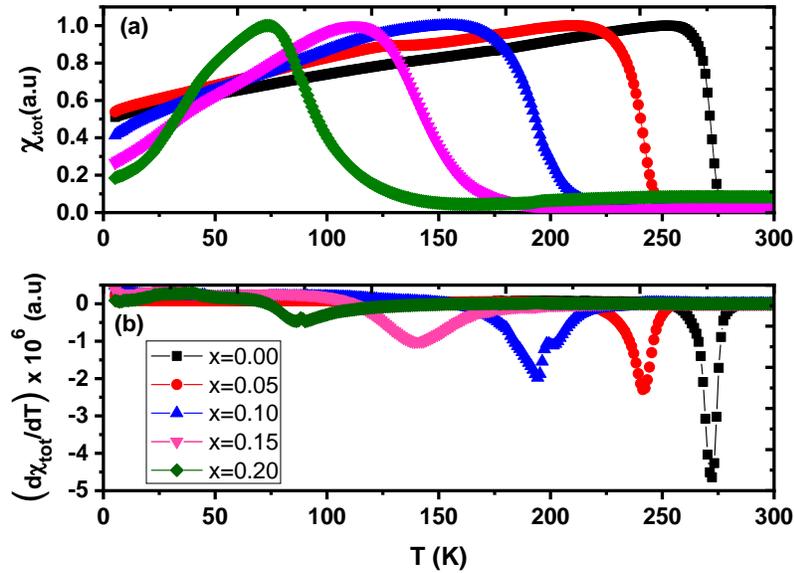

*Figure 5: (a) Normalized total ac-susceptibility for $Nd_{0.6}Sr_{0.4}Mn_{1-x}Ni_xO_3$ (0≤x≤0.2) polycrystalline, (b) their first derivative.*

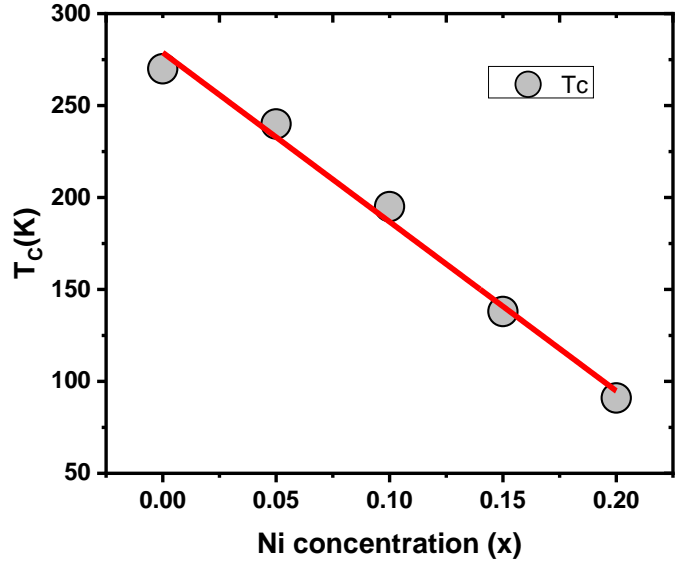

*Figure 6: Effect of Ni substitutions on critical temperature in $Nd_{0.6}Sr_{0.4}Mn_{1-x}Ni_xO_3$ compounds.*

### *3.3 The Modified Arrott plot (MAP)*

The conventional method to determine the critical exponents and transition temperature involves the use of Arrott plots or the modified Arrott plots along with Kouvel-Fischer plots [9, 10]. In the Arrott plot; the analysis is based on the mean field theory expansion of the free energy. The magnetization isotherms are presented as $M^2\ vs. H/M$. At relatively high fields (few *kOe*), the analysis leads to a set of parallel straight lines near T$_C$. At Tc, the magnetic isotherm passes through the origin. Deviations from linearity may persists at the high-field regions, hence reflecting deviations for the mean-field model. Banerjee suggested that the presence of curvature may indication of a second-order phase transition [25].

To accommodate these deviations, one may resort to critical behavior analysis using the modified Arrott-plot (MAP) techniques. This is based on Arrott-Noakes equation of state that holds well in the asymptotic critical region [9].

$$(H/M)^{1/\gamma} = (T - T_C)/T_1 + (M/M_1)^{1/\beta} \qquad (1)$$

Here, $T_1$ and $M_1$ are constants; $\gamma$ and $\beta$ are the critical exponents of the phase transition.

The critical behavior of phase transition has been extensively studied using various theoretical models for the magnetic spin configurations. The commonly used models are: 3D-Heisenberg ($\gamma = 1.336$ and $\beta = 0.365$), the tricritical mean-field ($\gamma = 1$ and $\beta = 0.25$), the 3D-Ising ($\gamma = 1.24$ and $\beta = 0.325$) and the 3D-XY ($\gamma = 1.316$ and $\beta = 0.345$). The third critical exponent $\delta$ can be evaluated using the magnetization isotherm at the transition temperature Tc. It can also be calculated from the Widom scaling relation $\delta = 1 + \frac{\gamma}{\beta}$ [26-27].

The modified Arrott analysis is greatly improved using Kouvel-Fisher plots in order to obtain reliable experimental values of the critical exponents.

The magnetization isotherms for all prepared samples are measured in the temperature range 4-300K and in an applied magnetic field 0-5 Tesla. A representative of these measurements is shown in Fig. 7 for temperatures 100K to just above Tc for each sample. The figures revealed a reduction in the saturated magnetization *Ms* with increasing Ni concentration. Moreover, the high field susceptibility, as obtained from the slope of the magnetization at high fields (~4Tesla) is increasing with increasing Ni contents. This also indicates that samples with high Ni-contents needs higher applied magnetic field to reach saturation. This also indicates strengthening of the antiferromagnetic interaction with increasing Ni-contents.

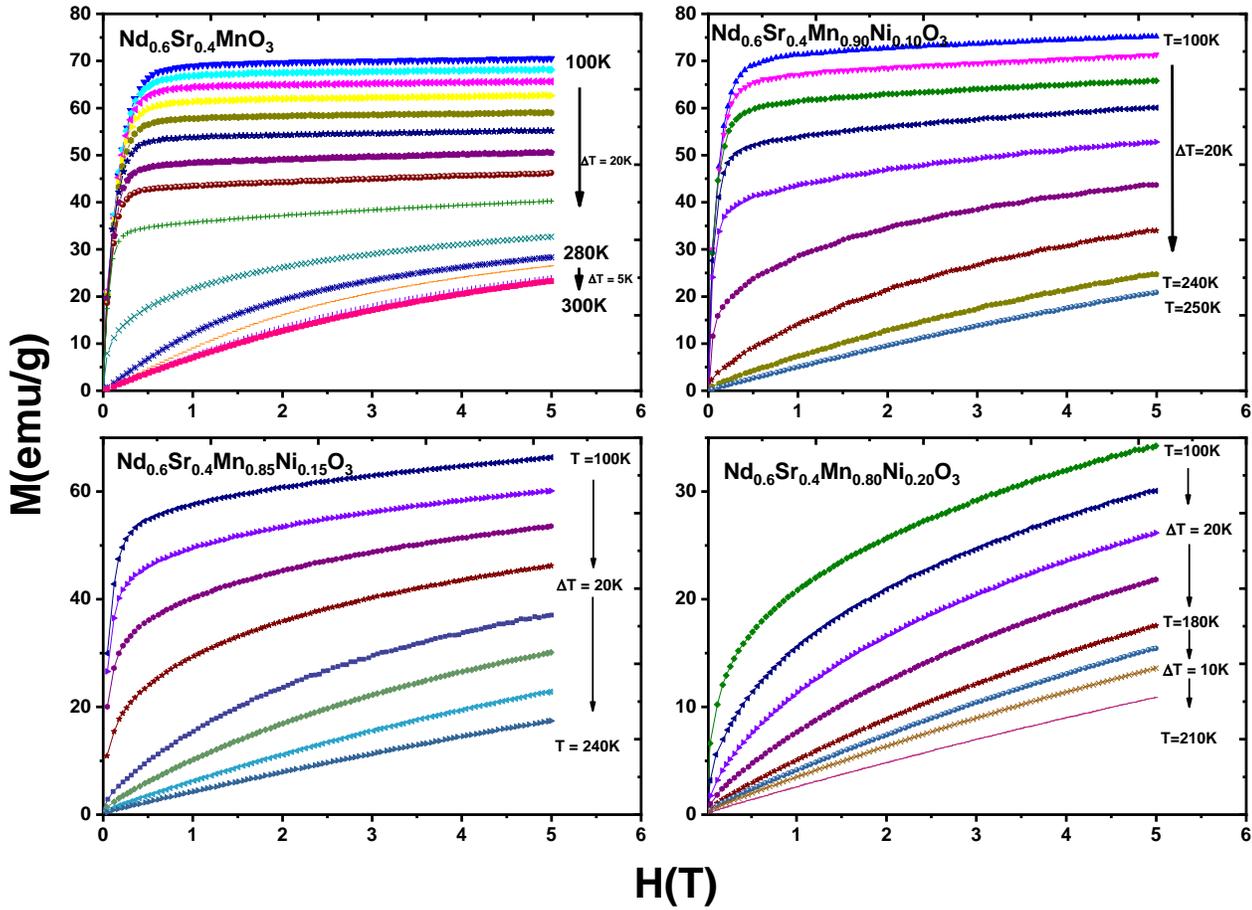

*Fig. 7: Magnetic isotherms for $Nd_{0.6}Sr_{0.4}Mn_{1-x}Ni_xO_3$ above 100K.*

The magnetic isotherms are represented as $M^{1/\beta}$ vs. $(H/M)^{1/\gamma}$ plots at different temperatures for x = 0.00 using four different models (Fig. 8): the 3D-Heisenberg model, the tricritical mean-field models, the 3D-Ising model, and the 3D-XY model. The theoretical values of $\beta, \gamma$ and $\delta$ for each model are given for example in reference [26-27].

Based on these curves (Fig. 8), all models render quasi-straight lines and nearly parallel at high field (~3Tesal). The slope of these lines relative to the slope at Tc (RS) is used to determine which model closely represents the magnetic state at hand. The variations of RS values -for the various models- with temperature are presented in figure 9. The most adequate model should be the one with RS curve close to RS=1 [28]. Therefore, one can infer from Fig. 9 that the 3D-Heisenberg model is the nearest to the horizontal line at RS = 1 for $Nd_{0.6}Sr_{0.4}MnO_3$ sample (x = 0.00), while for x = 0.05, the sample behaves like mean field model (Fig. 9).

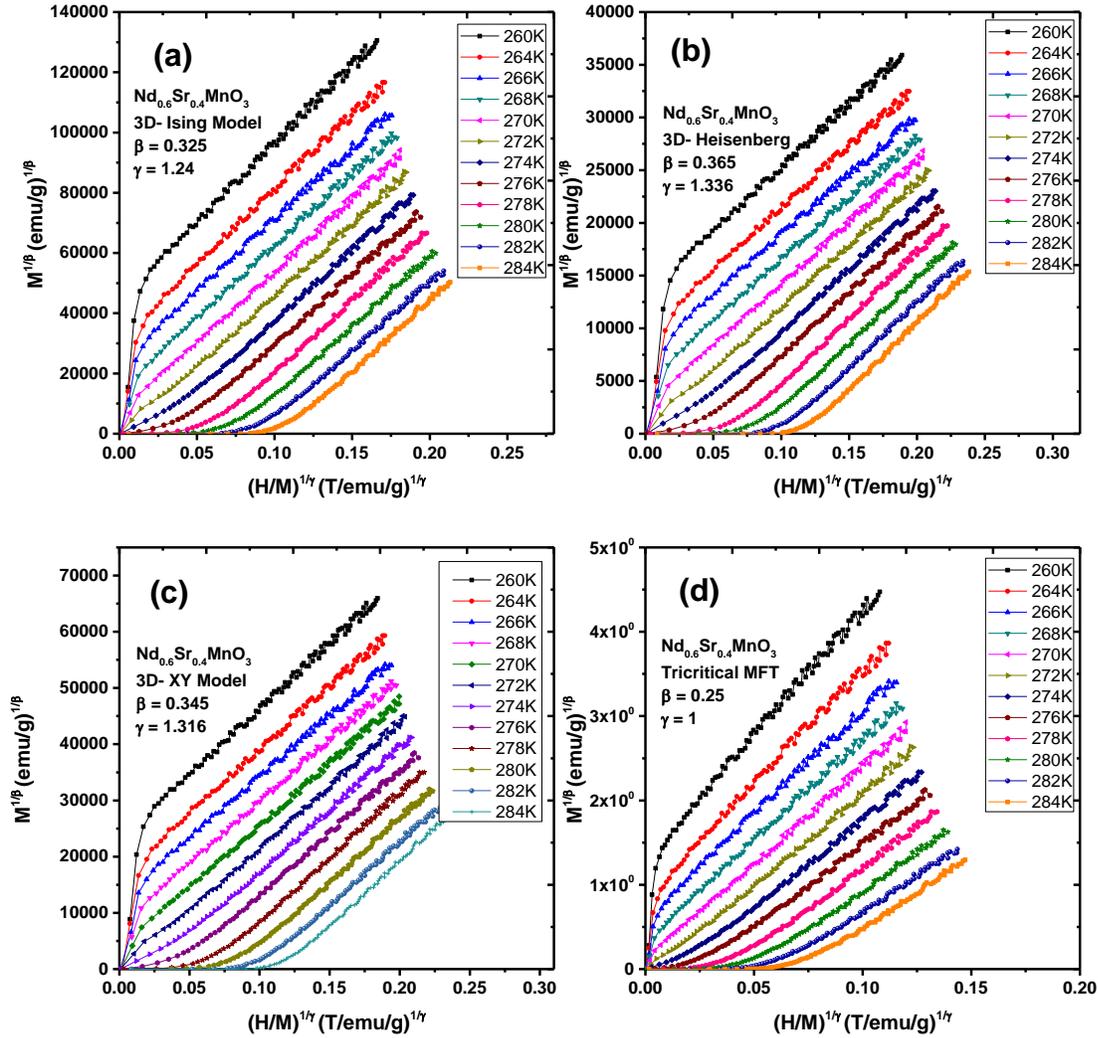

*Figure 8: Modified Arrott-plots for $Nd_{0.6}Sr_{0.4}MnO_3$ compound using different models; (a) 3D-Ising model, (b) 3D-Heisenberg model, (c) 3D-XY model, and (d) Tricritical mean field model.*

To estimate the experimental values of the critical exponents we used the high field (~4Tesla) magnetic isotherms to obtain the spontaneous magnetization $M_s(T)$ and the initial susceptibility $\chi_0^{-1}(T)$. The 3D-Heisenberg critical exponents for x=0.00 and the mean field critical exponents for the Ni-substituted samples x ≥ 0.05 samples have been used in the modified Arrott (MAP) plot representation. The variations of $M_s(T)$ and $\chi_0^{-1}(T)$ are presented in Fig. 10 (*a* and *b*) for x=0.00 and 0.05 respectively.

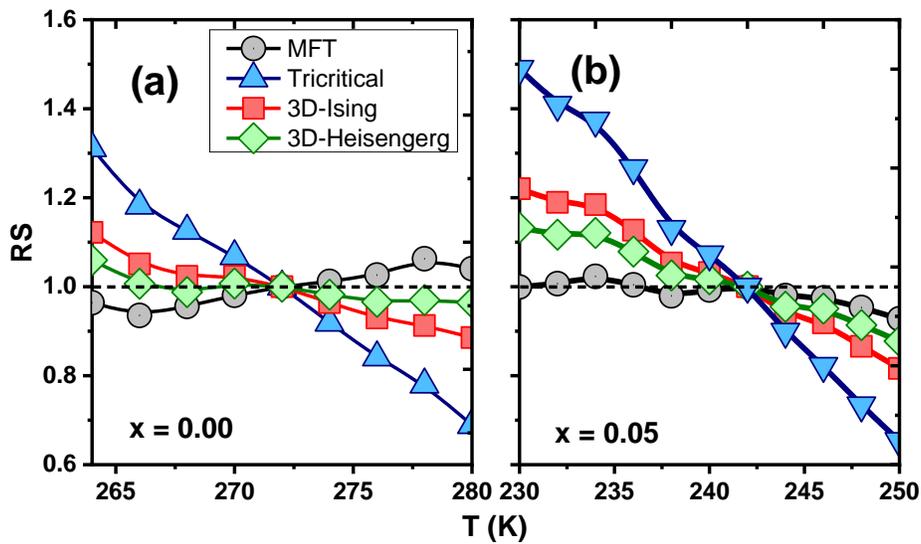

*Figure 9: Relative slope (RS) of (a) x = 0.00, and (b) x = 0.05 sample as a function of temperature defined as RS = S(T )/S(T$_C$).*

The values of critical exponents $\beta, and\ \gamma$ were obtained by fitting the experimental data with $M_S = m_0|\varepsilon|^\beta$; $when\ T < T_c$ and $\chi_0^{-1} = \frac{h_0}{m_0}|\varepsilon|^\gamma$; $when\ T > T_c$ where $\varepsilon = \frac{T-T_C}{T_C}$ is the reduced temperature, $m_0$, and $h_0$ are the critical amplitudes. The critical exponents obtained from the MAP along with Tc for all samples are shown in Table 4. The experimental value of the critical isotherm exponents δ for the Ni-free sample was obtained using the "critical" magnetic isotherm at Tc. The results (x=0.00) are presented in figure 11 as *ln(M)* vs *ln(H)* as shown in figure 11. The

$\delta$ value for the Ni substituted sample have been obtained using similar procedure. The results are given in Table 4, which clearly reveals that $\delta$ values are lower than the Ni-free sample.

*Table 4: Critical exponents $\beta, \gamma,$ and $\delta$ for $Nd_{0.6}Sr_{0.4}Mn_{1-x}Ni_xO_3$ compound.*

| X | Technique | $T_C(K)$ | $\beta$ | $\gamma$ | $\delta$ |
|---|---|---|---|---|---|
| 0.0 | MAP | 272.65 ± 0.54 | 0.314 ± 0.010 | 1.34 ± 0.07 | 4.7 |
| | K-F analysis | 273.62 ± 0.43 | 0.357 ±0.025 | 1.334±0.048 | |
| 0.05 | MAP | 246.07 ± 0.72 | 0.40 ± 0.01 | 1.05 ± 0.01 | 3.1 |
| | K-F analysis | 247.63 ± 0.38 | 0.49 ± 0.03 | 1.02 ± 0.01 | |
| 0.10 | MAP | 207.12 ± 0.35 | 0.47 ± 0.11 | 0.89 ± 0.05 | 2.9 |
| | K-F analysis | 206.45 ± 0.39 | 0.44 ± 0.15 | 0.82 ± 0.05 | |
| 0.15 | MAP | 152.12 ± 0.02 | 0.43 ± 0.04 | 0.99 ± 0.02 | 3.3 |
| | K-F analysis | 152.67 ± 0.19 | 0.36 ± 0.32 | 0.83 ± 0.06 | |
| 0.20 | MAP | 104.9 ± 0.24 | 0.40 ± 0.04 | 1.02 ± 0.14 | 3.1 |
| | K-F analysis | 103.64 ± 0.20 | 0.48 ± 0.02 | 1.03 ± 0.05 | |

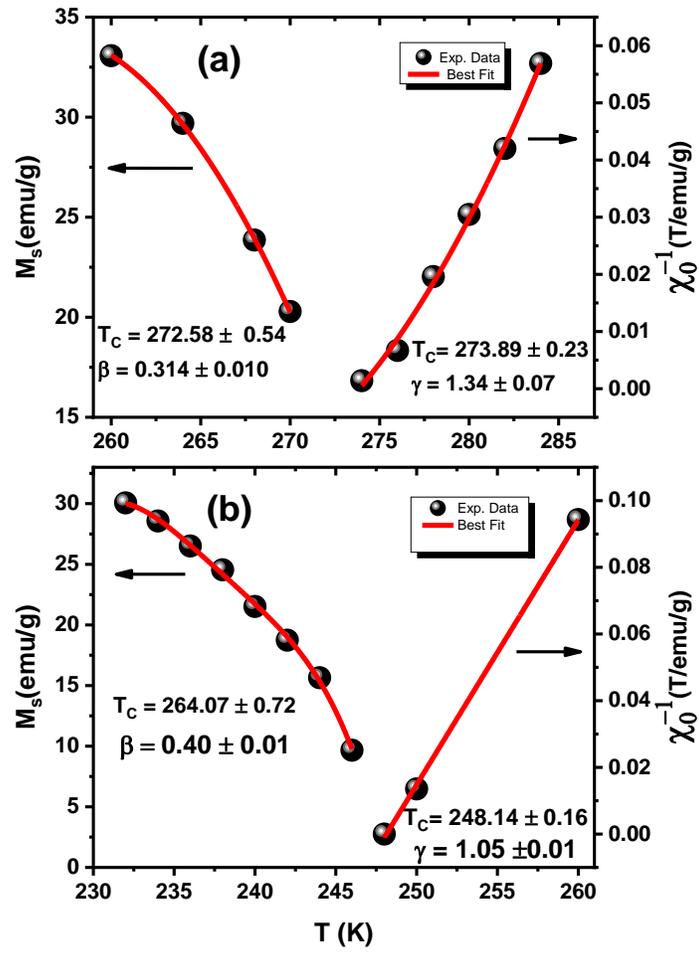

*Figure 10: Temperature dependence of Ms(T) and $\chi_0^{-1}$ determined from the modified Arrott plot for (a) x = 0.00, and (b) x = 0.05.*

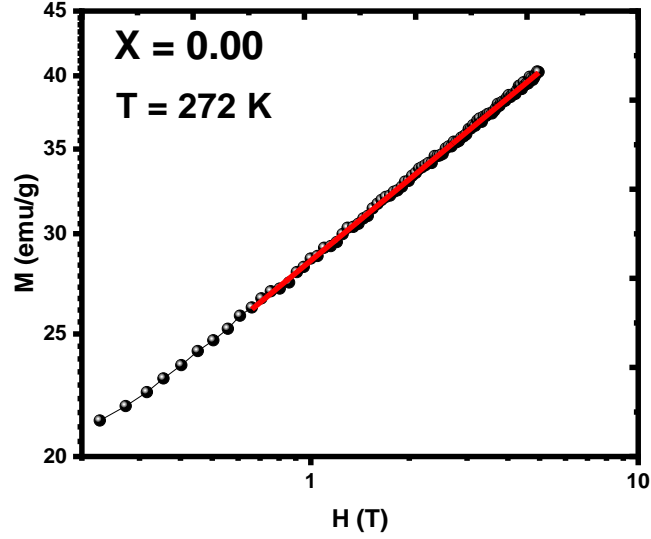

Figure 11: Critical isotherm at T=Tc in the Nd$_{0.6}$Sr$_{0.4}$MnO$_3$ compound in log-log scale.

### 3.4 The Kouvel–Fisher Plot

Kouvel–Fisher (KF) suggested a more precise method to determine the critical exponents [10]. The method is based on linearizing the two main order parameters; $M_s$ and $\chi_0^{-1}$ near Tc using the following relations [10]:

$$\frac{M_S(T)}{dM_S(T)/dT} = \frac{T-T_c}{\beta} \tag{2}$$

$$\frac{\chi_0^{-1}}{d\chi_0^{-1}/dT} = \frac{T-T_c}{\gamma} \tag{3}$$

The slopes are: $1/\beta$ and $1/\gamma$, respectively. The KF-plot is illustrated in Fig.12 for both x = 0.00 and x = 0.05 samples. For the Nd$_{0.6}$Sr$_{0.4}$MnO$_3$, the obtained critical exponents are ($\beta = 0.357 \pm 0.025$) and $\gamma = 1.334 \pm 0.048$ which are slightly different from the theoretical values used in the modified Arrott plots. For samples with x ≥ 0.05, the critical exponents' β and γ are very close to the mean field value. The critical exponents obtained from the MAP as well as the KF method along with $T_C$ are shown (for all samples) in Table 4 along with the calculate $\delta$ values [29].

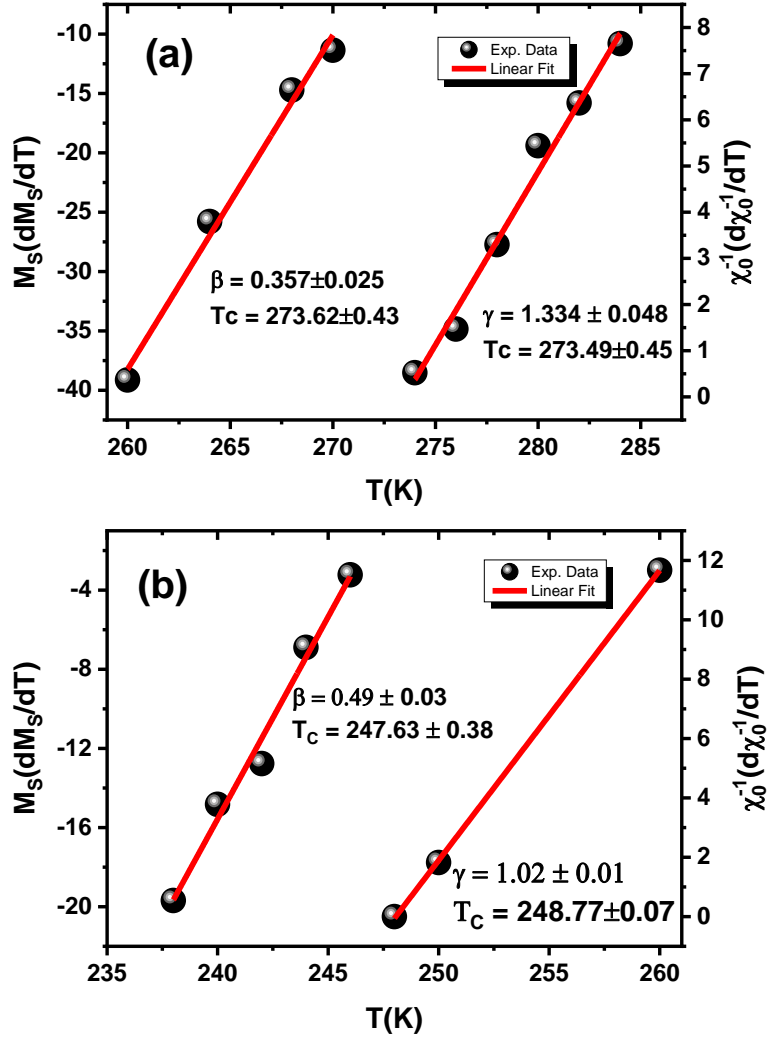

Figure 12: Kouvel-Fisher plots for Ms(T) and $\chi_0^{-1}$ plot of (a) x = 0.00, and (b) x = 0.05.

### 3.5 Scaling Analysis

Near the critical region the magnetic equation of state can be written as:

$$M(H, \varepsilon) = |\varepsilon|^\beta f_\pm \left(\frac{H}{|\varepsilon|^{\beta+\gamma}}\right) \qquad (4)$$

where $f_+$ is the state function above Tc (T >Tc) and $f_-$ is the state function below Tc (T < Tc), define ε both are polynomial functions [16]. The renormalized magnetization is defined as $m \equiv |\varepsilon|^{-\beta} M(H, \varepsilon)$, and renormalized field as $h \equiv H|\varepsilon|^{-(\beta+\gamma)}$. This means, Eq. 4 can be rewritten as

$$m = f_{\pm}(h) \tag{5}$$

For the correct values of the critical exponents, $\beta, \gamma$ and $\delta$ ; Eq. 5 will generate two distinct curves in $M/|\varepsilon|^{\beta}$ vs. $H/|\varepsilon|^{(\beta+\gamma)}$ representation: one curve for T above $T_C$ and another below $T_C$ [28]. Both curves approach each other asymptotically at T=Tc. This is an important criterion for critical regime. The scaled data is represented in Fig.13 a and b for x=0.0 and 0.05 respectively. All data at various temperatures fall on two distinct curves; one above $Tc$ and the other below $Tc$. Both curves merge at T=Tc. These findings confirm that the obtained values of the critical exponents and critical temperatures are reliable and consistent with the scaling hypothesis.

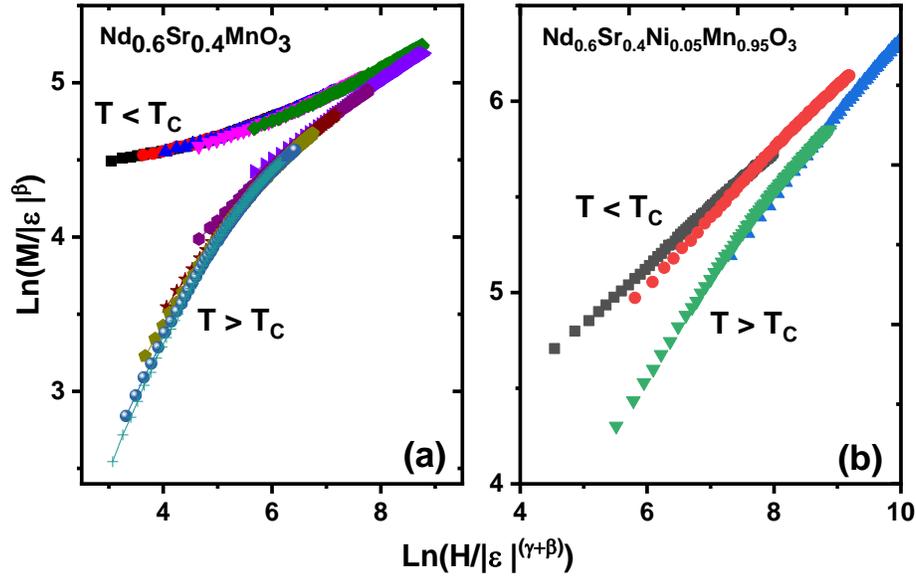

*Figure 13: Scaling plots indicating two universal curves below and above Tc for x=0.00 and x=0.05.*

**CONCLUSION**

The XRD patterns of $Nd_{0.6}Sr_{0.4}Mn_{1-x}Ni_xO_3$ with $0.00 \leq x \leq 0.2$ revealed that the major phase in all samples is in the orthorhombic structure with space group *Pnma*. The lattice parameters and the unit cell volume are found to decrease linearly with increasing Ni concentration. Moreover, the angle *O2-Mn-O2* decreases monotonically with increasing Ni concentration. The critical properties of these perovskite manganites have been studied using the dc magnetization near Tc, using the modified Arrott plot, Kouvel-Fisher method, critical isotherm and scaling analysis. The critical exponents for x = 0.00 sample closely follow a 3D Heisenberg model, while the Ni-substituted samples tend to be mean field like model.

The experimental results showed a second-order phase transition. The obtained critical exponents are found to produce a universal scaling of the magnetization data that falls into two distinct curves below and above the critical temperature. This confirms the reliability of the obtained critical exponents and Tc.

**ACKNOWLEDGMENTS**

The project has been supported by King Fahd University of Petroleum and Minerals DSR project No. IN121002. The support of the Center of Research Excellence in Nanotechnology (CENT) at KFUPM is gratefully appreciated.